\documentclass[12pt,prb]{revtex4}% Physical Review B

\usepackage{graphicx,epsfig}% Include figure files
\usepackage{dcolumn}% Align table columns on decimal point
\usepackage{bm}% bold math
\usepackage{setspace}

\newcommand{\mrd}{\hspace{0.6mm}}
\begin{document}

\preprint{}

\title{Quantum transport in semiconductor quantum dot superlattices: electron-phonon resonances and polaron effects}% Force line breaks with \\

  \author{Nenad Vukmirovi\'c\footnote{present address: Computational Research Division, Lawrence Berkeley National Laboratory, Berkeley, California 94720, USA, e-mail: NVukmirovic@lbl.gov}} 
   \author{Zoran Ikoni\'c}
  \author{Dragan Indjin}
 \email{d.indjin@leeds.ac.uk}
 \author{Paul Harrison}
 \affiliation{
 School of Electronic and Electrical Engineering,
 University of Leeds, Leeds LS2 9JT, United Kingdom
 }

 \singlespacing

\date{\today}% It is always \today, today,

\begin{abstract}
Electron transport in periodic quantum dot arrays in the presence of interactions with phonons was investigated using the formalism of nonequilibrium Green's functions. The self-consistent Born approximation was used to model the self-energies. Its validity was checked by comparison with the results obtained by direct diagonalization of the Hamiltonian of interacting electrons and longitudinal optical phonons.
The nature of charge transport at electron -- phonon resonances was investigated in detail and contributions from scattering and coherent tunneling to the current were identified. It was found that at larger values of the structure period the main peak in the current -- field characteristics exhibits a doublet structure which was shown to be a transport signature of polaron effects. At smaller values of the period, electron -- phonon resonances cause multiple peaks in the characteristics. A phenomenological model for treatment of nonuniformities of a realistic quantum dot ensemble was also introduced to estimate the influence of nonuniformities on current -- field characteristics.
\end{abstract}

\maketitle

\section{Introduction}

Semiconductor quantum dot superlattices are attracting increasing research attention due to their possible applications in a variety of devices. For example, they have the potential to increase solar energy conversion efficiency;~\cite{phe14-115} most recently intermediate-band solar cells based on them have been demonstrated.~\cite{prl97-247701} On the other hand, they are expected to have an improved thermoelectric figure of merit~\cite{apl82-415,sci297-2229} compared to bulk materials, paving the way for thermoelectric devices with improved performance. It is also expected~\cite{IEEEJQE33-1170,IEEEJSTQE6-491,pss(a)202-987} that quantum cascade lasers based on quantum dot superlattices should have superior performance compared to existing quantum well superlattice based quantum cascade lasers. Therefore, there is a significant interest in investigating the carrier transport in quantum dot superlattices, which is essential for understanding the performance of most of the devices mentioned.

In this paper, electron transport through periodic arrays of vertically stacked semiconductor quantum dots in the presence of the electron -- phonon interaction, will be investigated. Special emphasis will be put into the electron -- phonon resonances that occur when $nV_\mathrm{F}=mE_\mathrm{LO}$, where $n$ and $m$ are integers, $E_\mathrm{LO}$ the LO phonon energy, and $V_\mathrm{F}$ the potential drop over one period due to electric field. Resonances of this kind have been investigated in quantum well superlattices in an external axial magnetic field, that provides lateral confinement and causes a discrete electronic spectrum, similar to the one in quantum dots. It has been predicted~\cite{PhysRevB.43.9336,PhysRevB.59.8152,PhysRevB.53.7937,PhysRevB.56.15827} that inelastic optical phonon resonances occur whenever $nE_C+pV_F+qE_{LO}=0$ (Stark-cyclotron-phonon resonance), and elastic resonances (Stark-cyclotron) when $nE_C+pV_F=0$ (where $E_C$ is the cyclotron energy corresponding to the energy separation of Landau levels and $n$, $p$ and $q$ are integers). The observation of Stark-cyclotron resonances was reported in Ref.~\onlinecite{PhysRevLett.76.3618}. A special case of both of these, so called Stark-magneto-phonon resonances that occur when $nE_C=pV_F=qE_{LO}$ was recently measured, as reported in Refs.~\onlinecite{apl88-052111}~and~\onlinecite{prb74-121306}. 
Several theoretical studies have also addressed the transport through a few quantum dots in the presence of the electron -- phonon interaction. Phonon-assisted transport through a double quantum dot coupled to electric contacts was theoretically investigated in Ref.~\onlinecite{prb73-115338}, where a weak LO phonon interaction strength was assumed. The effect of a strong electron -- phonon interaction on the transport through a single quantum dot was studied in Ref.~\onlinecite{prb66-085311}.

The organization of the paper is as follows. The theoretical framework used will be presented in Sec.~\ref{Sec:chap9theory}, which will be additionally justified in Sec.~\ref{Sec:chap9SCBA}. The overall characteristics of the current -- field curves will be discussed in Sec.~\ref{Sec:chap9curr}, with special emphasis on the main peak in the curves in Sec.~\ref{Sec:chap9mainpeak} and other peaks in Sec.~\ref{Sec:chap9otherres}. The influence of nonuniformities will be estimated in Sec.~\ref{sec:chap9secNON}.

\section{Theoretical framework}\label{Sec:chap9theory}
The formalism of nonequilibrium Green's functions~\cite{Haug,pr357-1} was used to evaluate the current in the steady state through an array of identical quantum dots.
The central quantities are expectation values of product or anticommutators of electron creation and annihilation operators at different times, such as the retarded Green's function
\begin{equation}
G^R_{\alpha\beta}(t_1,t_2)=-i\Theta(t_1-t_2)\langle \{\hat{a}_\alpha(t_1),\hat{a}_\beta^+(t_2)\}\rangle,
\end{equation}
the advanced Green's function
\begin{equation}
G^A_{\alpha\beta}(t_1,t_2)=i\Theta(t_2-t_1)\langle \{\hat{a}_\alpha(t_1),\hat{a}_\beta^+(t_2)\}\rangle=G^R_{\beta\alpha}(t_2,t_1)^*,
\end{equation}
and the lesser Green's function
\begin{equation}\label{eq:chap2less}
G^<_{\alpha\beta}(t_1,t_2)=i\langle \hat{a}_\beta^+(t_2)\hat{a}_\alpha(t_1)\rangle.
\end{equation}
As can be seen from (\ref{eq:chap2less}), the lesser function at equal times represents populations and coherences of the states, in terms of which other relevant physical quantities can be expressed. In the steady state of the system, Green's functions depend only on the difference of their time arguments. One can therefore define the Fourier transform of all these quantities as
\begin{equation}
F(E)=\int \mathrm{d}(t_1-t_2) e^{iE(t_1-t_2)/\hbar}F(t_1-t_2).
\end{equation} 
In order to find the retarded and the lesser functions, one has to solve their dynamical equations.
These satisfy the Dyson equation
\begin{equation}\label{eq:chap2Dysonspec}
\sum_\gamma\left[E\delta_{\alpha\gamma}-\left(H_{\alpha\gamma}+\Sigma_{\alpha\gamma}^R(E)\right)\right]G_{\gamma\beta}^R(E)=\delta_{\alpha\beta}
\end{equation}
and the Keldysh relation 
\begin{equation}\label{eq:chap2Keldyshspec}
G_{\alpha\beta}^<(E)=\sum_{\gamma\delta}G_{\alpha\gamma}^R(E)\Sigma_{\gamma\delta}^<(E)G_{\delta\beta}^A(E).
\end{equation}
The relation between the retarded and advanced function in the energy domain is  
\begin{equation}
G_{\alpha\beta}^R(E)=G_{\beta\alpha}^A(E)^*.
\end{equation}
The Hamiltonian $\hat{H}$  contains the kinetic energy of the electron and single particle potential, while all other interactions are contained in the self-energy $\Sigma$.

As a first step in the application of the formalism to a given physical system, one has to choose the basis of states to represent the Green's functions. Here, a basis of states localized mainly to one period is chosen. Such a choice enables one to make   an approximation where interactions with nearest neighbors only are considered. Additionally, such a basis gives an excellent insight into the carrier transport in real space. Due to the periodicity of the structure, the basis states are labelled as $(\nu,n)$, where $\nu$ is the index of the state assigned to period $n$ in ascending order of energies. In the case when only ground states are important, the first index can be suppressed. The basis states are calculated as follows.

The electronic miniband structure of a quantum dot superlattice is solved using the eight band ${\bm k}\cdot {\bm p}$ method with the strain distribution taken into account via continuum elasticity theory, as described in more detail in Ref.~\onlinecite{sst21-1098}. 
As a result of this step one obtains the quantum dot superlattice eight component spinors $|\Psi_{\nu K_z}({\bm r})\rangle$, satisfying the Bloch condition
\begin{equation}
|\Psi_{\nu K_z}({\bm r}+L_z{\bm e}_z)\rangle=e^{iK_zL_z}|\Psi_{\nu K_z}({\bm r})\rangle,
\end{equation}
where $\nu$ is the miniband index, $L_z$ the period of the structure, and $K_z$ the superlattice wave vector. The phase of the spinors was fixed by imposing the condition that the value of the dominant spinor component at a particular point in space is real and positive.

The spinors obtained are then used to construct Wannier states that are localized to a certain period. The Wannier state originating from miniband $\nu$, localized to period $n$ is given by
\begin{equation}
|\Psi_{\nu n}\rangle=\frac{L_z}{2\pi}\int_{-\pi/L_z}^{\pi/L_z}\mathrm{d}K_ze^{-inK_zL_z}|\Psi_{\nu K_z}\rangle.
\end{equation}
In order to obtain states with even better localization (i.e. the states with the probability of finding the electron in period $n$ being closer to 1), the eigenvalue problem of the operator of the $z$-coordinate is solved in the manifold of states spanned by $|\Psi_{\nu n}\rangle$, $n\in\{-N,\ldots,N\}$. The $(N+1)$-th eigenvector then corresponds to the basis state $(\nu,0)$. The states $(\nu,n)$, when $n\ne 0$ are then obtained by making a translation in real space by $nL_z$. Since the eigenstates of the position operator in the total vector space of the system are fully localized delta functions, it is expected that the procedure described, performed in a limited subspace, yields states with improved degree of localization. The actual calculation, where $N$ was overcautiously set to 10, verified this expectation. Additional convenience of this basis is the fact that the external potential operator $|e|F\hat{z}$ (where $F$ is the electric field) is diagonal when the basis is restricted to the states with $\nu=1$. However, when the states with $\nu>1$ are included, this is no longer the case.

Once the basis of states is chosen, one can proceed to calculate the relevant Green's functions represented in that basis and afterwards the current in the structure. In the steady state of the system, one obtains an algebraic system of equations for Green's functions in the energy domain, containing the Dyson equation~(\ref{eq:chap2Dysonspec}), the Keldysh relation~(\ref{eq:chap2Keldyshspec}), and the expressions for self-energies. The system of equations is closed by imposing the periodic condition for all Green's functions and self-energies
\begin{equation}
G_{(\nu,n),(\mu,m)}(E)=G_{(\nu,n+1),(\mu,m+1)}(E+V_F),
\end{equation}
and introducing   an approximation by considering only the Green's functions and self-energies with $|n-m|\le K$.

The interactions with phonons considered in this work are polar coupling to optical phonons and deformation potential coupling to acoustic phonons, as it is known that other electron -- phonon interaction mechanisms, such as deformation potential coupling to optical phonons and piezoelectric coupling to acoustic phonons are less important.~\cite{Mahan} As it is thought that the influence of phonon confinement 
% on scattering rates 
is not so important in AlGaAs/GaAs and InGaAs/GaAs nanostructures,~\cite{apl65-469,prb45-8756}
bulk phonon modes are assumed. 
The Fr\"{o}lich interaction Hamiltonian describing polar coupling to optical phonons is then given by~\cite{Harrison,Mahan}
\begin{equation}\label{eq:chap2jed2}
\hat{H}_{e-ph}=\sum_{ij{\bm q}}M_{ij}({\bm q})\hat{a}_i^+\hat{a}_j\left(\hat{b}_{\bm q}+\hat{b}_{-\bm q}^+\right),
\end{equation}
where $\hat{b}_{{\bm q}}$ and $\hat{b}^+_{{\bm q}}$ are phonon annihilation and creation operators, $M_{ij}({\bm q})=\alpha({\bm q})F_{ij}({\bm q})$,
\begin{equation}\label{eq:jed3}
\alpha({\bm q})=\frac{1}{q}\sqrt{\frac{e^2E_{\textrm{LO}}}{2V}
\left(\frac{1}{\varepsilon_{\infty}}-\frac{1}{\varepsilon_{st}}\right)},
\end{equation}
$V$ is the volume of the box used for discretization of ${\bm q}$ vectors, $F_{ij}({\bm q})$
is the electron -- phonon interaction form factor,~\cite{Harrison} and $\varepsilon_{\infty}$ and $\varepsilon_{st}$ are high frequency and static dielectric constants, respectively. Optical phonons are nearly dispersionless and for simplicity, a constant LO phonon energy $E_\mathrm{LO}$ is assumed.

The Hamiltonian of the deformation potential interaction with acoustic phonons is given by the same formula (\ref{eq:chap2jed2}) except that in this case
\begin{equation}
\alpha({\bm q})=\sqrt{\frac{D_A^2\hbar q}{2\rho v_sV}},
\end{equation}
where $D_A$ is the acoustic deformation potential, $\rho$ the material density and $v_s$ the longitudinal sound velocity. To a very good approximation, a linear and isotropic acoustic phonon dispersion relation $\omega({\bm q})=v_sq$ can be assumed.

Self-energies are modelled using the self-consistent Born approximation (SCBA). Within the SCBA, self-energies due to the electron -- phonon interaction 
in the system with translational invariance~\cite{ap236-1} are given by the Fock term~\cite{Haug,Mahan,pr357-1}
\begin{eqnarray}\label{eq:chap2SigmaRspec}
\Sigma^R_{\alpha\beta}(E)=i\sum_{\gamma\delta,{\bm q}}M_{\beta\delta}({\bm q})^*M_{\alpha\gamma}({\bm q})
\frac{1}{2\pi}\int\mathrm{d}E'
\left[
G^R_{\gamma\delta}(E-E')D^R(E')+\right.\\ \left.+
G^<_{\gamma\delta}(E-E')D^R(E')+
G^R_{\gamma\delta}(E-E')D^<(E')
\right],\nonumber
\end{eqnarray}
\begin{equation}\label{eq:chap2SigmaLspec}
\Sigma^<_{\alpha\beta}(E)=i\sum_{\gamma\delta,{\bm q}}M_{\beta\delta}({\bm q})^*M_{\alpha\gamma}({\bm q})
\frac{1}{2\pi}\int\mathrm{d}E'
G^<_{\gamma\delta}(E-E')D^<(E')
.
\end{equation}
 In the systems lacking translational invariance, such as a single quantum dot investigated in Sec.~\ref{Sec:chap9SCBA}, there is an additional contribution to $\Sigma^R$ from the Hartree term (see for example Ref.~\onlinecite{prb73-115338} for the explicit expression). In the limit of low carrier density ($G^<=0$) investigated in Sec.~\ref{Sec:chap9SCBA}, this term vanishes and therefore it was not considered.
The anharmonic decay of LO phonons, which is known to be important for the proper description of relaxation processes in quantum dots,~\cite{prb59-5069} was taken into account by adding an exponentially decaying term~\cite{prl85-1516} to the free phonon Green's functions in the time domain. The phonon Green's functions in the energy domain are then given by
\begin{equation}
D^R(E)=\frac{1}{E-E_{{LO}}+i\Gamma}-\frac{1}{E+E_{{LO}}+i\Gamma},
\end{equation}
\begin{equation}
D^<(E)=-i\left[(N_{LO}+1)\frac{2\Gamma}{(E+E_{LO})^2+\Gamma^2}+
N_{LO}\frac{2\Gamma}{(E-E_{LO})^2+\Gamma^2}
\right],
\end{equation}
where $\Gamma$ is the LO phonon linewidth determined by its anharmonic decay rate and $N_\mathrm{LO}$ is the phonon occupation number 
\begin{equation}
N_\mathrm{LO}=\frac{1}{e^{\frac{\hbar\omega_\mathrm{LO}}{k_\mathrm{B}T}}-1}.
\end{equation}

Self-energy terms due to the interaction with acoustic phonons are given by the formulas which have the same form as in the case of LO phonons. These can be simplified to avoid a demanding integration in the energy domain, assuming acoustic phonons are stable. They then read~\cite{pr357-1}
\begin{eqnarray}
\Sigma_{\alpha\beta}^R(E)=
\sum_{\gamma\delta,{\bm q}}
M^*_{\beta\delta}({\bm q})M_{\alpha\gamma}({\bm q})
&\left[
(N_{\bm q}+1)G_{\gamma\delta}^R(E-E_{\bm q})+
 N_{\bm q}   G_{\gamma\delta}^R(E+E_{\bm q})+\right.\nonumber\\
&\left.+\frac{1}{2}G_{\gamma\delta}^<(E-E_{\bm q})-
\frac{1}{2}G_{\gamma\delta}^<(E+E_{\bm q})
\right],
\end{eqnarray}
\begin{equation}
\Sigma_{\alpha\beta}^<(E)=
\sum_{\gamma\delta,{\bm q}}
M^*_{\beta\delta}({\bm q})M_{\alpha\gamma}({\bm q})
\left[
N_{\bm q}    G_{\gamma\delta}^<(E-E_{\bm q})+
(N_{\bm q}+1)G_{\gamma\delta}^<(E+E_{\bm q})
\right],
\end{equation}
where $E_{\bm q}$ is the energy of an acoustic phonon and $N_{\bm q}$ is the acoustic phonon occupation number. The principal value integrals appearing in the expression for the retarded self-energy have been neglected, as is often done in the literature.~\cite{PhysRevB.66.245314}

An additional self-energy term representing the nonuniformity of quantum dots can also be included, as described in Sec.~\ref{sec:chap9secNON}.

The justification of application of the SCBA to electron -- LO phonon interaction self-energies in the system studied here will be given in Sec.~\ref{Sec:chap9SCBA}. In the expressions for self-energies, only the electron -- phonon interaction form factors between states with $|n-m|\le K$ are assumed to be nonvanishing.

The interest here will be in the limit of low doping and carrier densities where the interaction with ionized impurities and electron -- electron interaction can be neglected, and there is no formation of electric field domains. In this region, current depends linearly on the number of carriers. Therefore, the values of current presented  have been normalized by dividing it by the total occupancy of states in one quantum dot.

The system of algebraic equations for Green's functions and self-energies was solved in a manner that is now described. 
When the current -- field characteristic is calculated, i.e. when the same calculation is performed for different values of the electric field, the results obtained for the previous value of the field can be used as an initial guess. Otherwise, an initial guess for the lesser Green's functions is taken in the form $
G^<_{\alpha\beta}(E)=2\pi i\mrd g(E-E_\alpha,\sigma)\mrd n_\alpha\delta_{\alpha\beta},
$
where $g$ is the Gaussian function, and $n_\alpha$ is the initial guess for expected values of state populations given by the thermal distribution of carriers. The initial guess for the retarded Green's function is obtained from the self-consistent solution of Eqs. (\ref{eq:chap2Dysonspec}) and (\ref{eq:chap2SigmaRspec}) where the terms with lesser electron Green's function have not been included. After the initial guess has been established, retarded and lesser self-energies are calculated. Next, the retarded Green's functions are calculated from the Dyson equation by solving the appropriate system of linear equations. Finally the lesser Green's function is calculated from the Keldysh relation. These three steps constitute one iteration of the self-consistent procedure which is repeated until convergence is achieved.  In order to improve the stability of the self-consistent procedure, the lesser and retarded functions for the next iteration are calculated from their average value in the previous two iterations, as is usually done in self-consistent calculations. After each iteration the lesser Green's functions are adjusted to enable the total number of particles to be equal to a given predefined value.

It should be mentioned that due to the assumption of dispersionless LO phonon modes, the integral in the expression for self-energy does not depend on ${\bm q}$. Therefore, the terms $M_{\alpha\beta\gamma\delta}=\sum_\mathrm{q}M^*_{\beta\delta}({\bm q})M_{\alpha\gamma}({\bm q})$ can be calculated only once before the self-consistent procedure, rather than in each iteration.
When the self-energies due to the interaction with acoustic phonons are 
concerned, due to the assumption of isotropic dispersion relation these take the form
$
\int{d}^3{\bm q}M^*_{\beta\delta}({\bm q})M_{\alpha\gamma}({\bm q})f(|q|).
$
The integral over spherical coordinates $\theta$ and $\varphi$ for each $|q|$ can therefore be calculated before the self-consistent procedure. However, the integral over $|q|$ must be calculated in each iteration.

The populations of the energy levels and coherences between states can finally be calculated by performing an integration of lesser Green's functions over the whole energy domain. The current through the structure can be calculated as described in Sec.~\ref{Sec:chap9curr}.

\section{Validation of the self-consistent Born approximation}\label{Sec:chap9SCBA}

The main approximation in the model described is the use of the SCBA, which therefore needs to be validated. Although the SCBA was widely used for modeling the electron transport in quantum well based superlattices~\cite{pr357-1} and quantum cascade structures,~\cite{PhysRevB.66.245314} it is not immediately apparent that it should be valid also for quantum dot superlattices.

The SCBA was used in Ref.~\onlinecite{prb73-115338} to describe the transport through two quantum dots coupled to contacts in the presence of the electron -- LO phonon interaction. The electron -- phonon interaction matrix elements $M_{\alpha\alpha\alpha\alpha}$ used in Ref.~\onlinecite{prb73-115338} were of the order $\sim 0.001\times E_{\mathrm{LO}}^2$, implying a weak interaction where the SCBA is fully justified, and it has been argued~\cite{prb73-115338} that polaron effects become important when $M_{\alpha\alpha\alpha\alpha}$ approaches $E_{\mathrm{LO}}^2$, which is expected to be the regime of strong electron -- phonon coupling, beyond the reach of SCBA.
In Ref.~\onlinecite{prb71-125327}, polaron relaxation in InGaAs quantum dots assisted by the presence of wetting layer states was treated within the SCBA (called random phase approximation therein) . It has been pointed out there that the SCBA is expected to be valid in the presence of continuum states provided by the wetting layer, which has been verified by a comparison with the first term in the cummulant expansion.~\cite{prb71-125327} On the other hand, in Ref.~\onlinecite{PhysRevB.62.7336} the problem of interaction of quantum dot carriers with dispersionless LO phonon modes was treated, and the conclusion was reached that the SCBA cannot reproduce the exact solution of the idealized model given by a series of delta functions at all temperatures. This is a consequence of the fact that the SCBA sums only a limited number of diagrams in the expansion, while a full summation is needed to reproduce delta functions.

One cannot conclude from the previous works just mentioned~\cite{prb73-115338,prb71-125327,PhysRevB.62.7336} whether the SCBA is a good approximation in the system considered here. Numerical calculation of the $M_{\alpha\alpha\alpha\alpha}$ matrix element (where $\alpha$ is the ground state) gives the value of $\sim 0.07\times E_\mathrm{LO}^2$, which is larger than the value used in Ref.~\onlinecite{prb73-115338} but still significantly smaller than $E_{\mathrm{LO}}^2$. The validity of the RPA in Ref.~\onlinecite{prb71-125327} was established for a quantum dot system in the presence of nearby wetting layer   states, while here the interest is mainly in transport through bound quantum dot states. In contrast to Ref.~\onlinecite{PhysRevB.62.7336} where a single quantum dot interacting with dispersionless LO phonons only is considered, other interactions are included in the system considered here, such as anharmonic terms leading to LO phonon decay, the interaction with acoustic phonons, as well as an additional term due to nonuniformity of the quantum dot ensemble.

In order to validate the use of the SCBA, it will be established here that for InAs/GaAs quantum dots, the polaron shift of the ground state, as well as the polaron splitting when the energy difference between the ground and first excited state is set to an LO phonon resonance, are accurately calculated in the SCBA. This gives confidence that the positions of the peaks of Green's functions are correct. The physical properties of the system depend not only on the positions of the peaks but also on their linewidths. One therefore has to establish that the linewidths originate from real interactions in the system, rather than from the effect described in Ref.~\onlinecite{PhysRevB.62.7336}. This will be done by showing that the calculated linewidths in the presence of acoustic phonons are significantly larger than the ones arising due to artificial broadening of the spectrum of electrons interacting with dispersionless LO phonons only.

\begin{figure}[htbp]
\vspace{1cm}
\centering
\epsfig{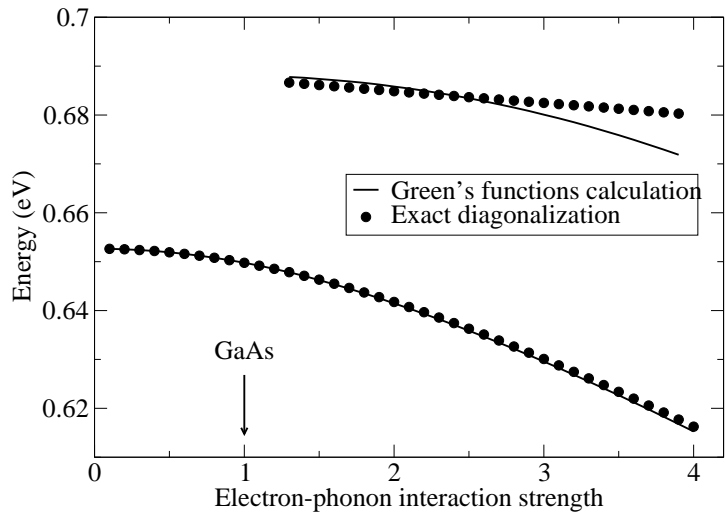}
 \caption{The dependence of the ground state energy of and its first phonon replica of a single InAs/GaAs quantum dot on the electron -- LO phonon interaction strength. A comparison of the results obtained by exact diagonalization of the electron -- LO phonon interaction Hamiltonian (circles) and by the Green's function calculation in the SCBA (full line) is given.}
  \label{fig:grpolfig}
\end{figure}

A comparison of the polaron shifts in the spectrum calculated by direct diagonalization of the Hamiltonian of electrons and LO phonons whose interaction is described by the Hamiltonian Eq.~(\ref{eq:chap2jed2}), and by the Green's functions method, for different electron -- phonon interaction strengths, is given in Fig.~\ref{fig:grpolfig}. The calculations were performed for a lens shaped single InAs/GaAs quantum dot of diameter 20~nm and height 5~nm, which is representative of typical self-assembled quantum dots obtained in experiments. The electronic structure of the quantum dot was calculated using the eight band strain dependent ${\bm k}\cdot{\bm p}$ model as described in more detail in Ref.~\onlinecite{sst21-1098}. The electronic states obtained that way were subsequently used as input for both calculations. The strength of the electron -- phonon interaction was artificially varied by multiplying the electron -- phonon interaction Hamiltonian by a constant whose value is given on the $x-$axis in Fig.~\ref{fig:grpolfig}.

\begin{figure}[htbp]
\vspace{1cm}
\centering
 \epsfig{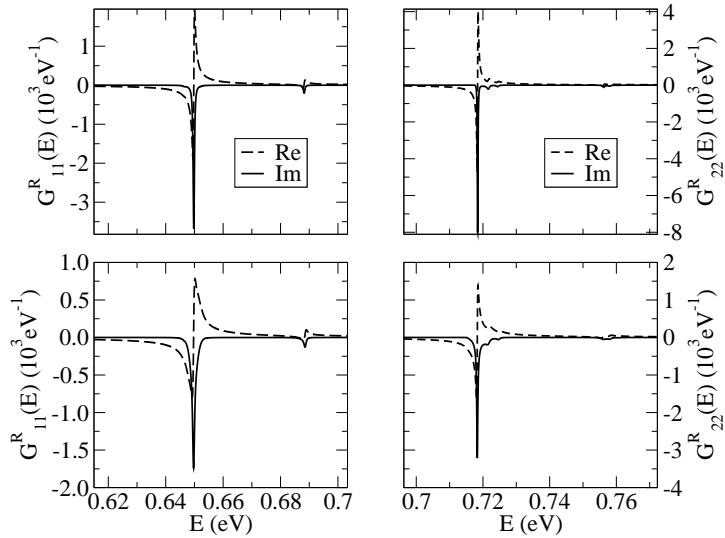}
 \caption{Real (dashed line) and imaginary (full line) part of the retarded Green's function of the ground state (left) and first excited state (right) in the case when the interaction with acoustic phonons is excluded (top) and included (bottom).}
  \label{fig:GR}
\end{figure}

In order to provide a fair comparison, in both calculations, only the ground and the pair of nearly degenerate first excited states were taken into account, and only the electron -- LO phonon interaction was considered. Direct diagonalization is performed using the method of Refs.~\onlinecite{PhysRevB.62.7336} and \onlinecite{prb73-115303}, where a unitary transformation on phonon modes is performed in such a way that only a few phonon modes remain coupled with electronic degrees of freedom, therefore enabling efficient diagonalization. The energies of the polaron states that contain a contribution from the purely electronic ground state of more than 10\% are represented by circles in Fig.~\ref{fig:grpolfig}. The Green's function calculation was performed by self-consistently iterating between Eqs. (\ref{eq:chap2Dysonspec}) and (\ref{eq:chap2SigmaRspec}) in the limit of low numbers of carriers (lesser electronic Green's functions set to zero), where a temperature of $T=77\mrd\mathrm{K}$ and LO phonon linewidth of $\Gamma=0.13\mrd\mathrm{meV}$ was assumed. A fully fair comparison would require $T=0$ and $\Gamma=0$, however the positions of the peaks weakly depend on $T$ and $\Gamma$ as they decrease from the values used to zero. In Fig.~\ref{fig:grpolfig} the positions of the maximum of the spectral function $A_{11}(E)=-2\mathrm{Im}G^{\mathrm{R}}_{11}(E)$, and its replica when its peak value is at least 10\% of the main maximum peak value are shown with a full line. The retarded Green's functions of the ground and first excited state are given in the top part of Fig.~\ref{fig:GR}. One can see from Fig.~\ref{fig:grpolfig} that excellent agreement for the polaron shift of the ground state obtained by the two methods is obtained throughout the whole interval of electron -- phonon interaction strengths investigated. On the other hand, for larger interaction strengths (say larger than 2.5) the positions of the replica start to differ. Further presentations will show that this replica is important for the description of carrier transport. Therefore, the conclusion arising from the results presented in Fig.~\ref{fig:grpolfig} is that the application of the SCBA can be expected to give reliable prediction of polaron shifts up to the electron -- phonon interaction strength being 2.5 times larger than the strength in the InAs/GaAs material system which is of central interest here. 

\begin{figure}[htbp]
\vspace{1cm}
\centering
 \epsfig{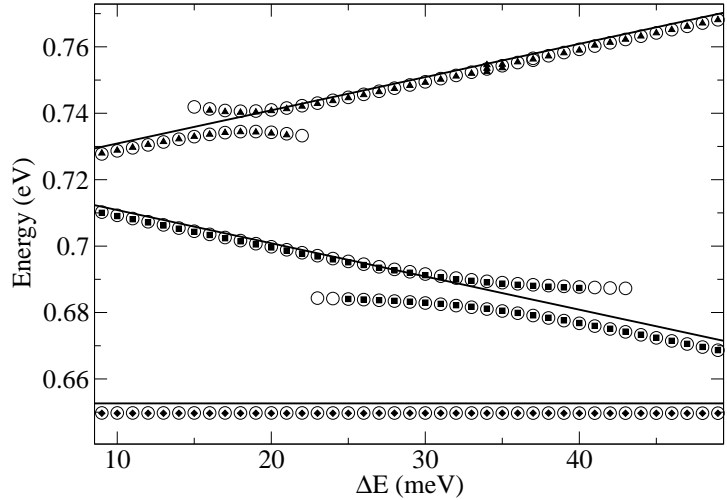}
 \caption{Dependence of the polaron energy levels obtained by direct diagonalization (circles) and the maxima of the spectral functions $A_{ii}(E)$ obtained in the SCBA (diamonds $i=1$, squares $i=2$ and triangles $i=3$) on the artificial shift $\Delta E$. The corresponding single-particle levels are shown by full lines.}
  \label{fig:chap9splitt}
\end{figure}

It is shown next that the SCBA also accurately predicts the amount of polaron splitting when two levels are at an LO phonon resonance. For that purpose, a numerical experiment is performed where the energies of the pair of first excited states are shifted in opposite directions by the same amount $\Delta E$, which is varied. The electron -- phonon interaction matrix elements are kept constant.  The polaron energy levels that contain a contribution from at least one of the electronic states larger than 10\% are shown by circles in Fig.~\ref{fig:chap9splitt}, while the maxima of the spectral functions $A_{ii}$ whose peak values are at least 10\% of the main peak value are represented by diamonds, squares and triangles, for $i=1$, $i=2$ and $i=3$, respectively. The results obtained by the SCBA are in excellent agreement  with the results obtained by direct diagonalization.

 The Green's functions of the two states when the interaction with acoustic phonons is included in the calculation are shown in the bottom part of Fig.~\ref{fig:GR}. In the previous case (no acoustic phonons, top part of Fig.~\ref{fig:GR}) the linewidth originated from the combined effect of artificial broadening due to limitations of the SCBA and from the finite phonon lifetime. Since much larger linewidths are obtained in this case, one can conclude that they do originate from the interactions in the system rather than from the artifacts of the SCBA.

\section{Transport in an ideal superlattice of quantum dots}\label{Sec:chap9curr}
The electron current through an array of quantum dots can be expressed in terms of the expectation value of the velocity operator as~\cite{prb73-245320,prb72-125347}
\begin{equation}
I=-\frac{|e|}{L}\left\langle\frac{\mathrm{d}\hat{Z}}{\mathrm{d}t}\right\rangle,
\end{equation}
where $L$ is the total length of the structure in the $z-$direction and $\hat{Z}$ is the coordinate operator of all the electrons in the system in the Heisenberg picture. 
From its equation of motion, using the definition of the lesser Green's function, by exploiting the periodicity of the structure, and bearing in mind that all interaction terms commute with $\hat{Z}$, as emphasized in Ref.~\onlinecite{prb73-245320}, it follows that
\begin{equation}\label{eq:chap9current}
I=-\frac{|e|}{L_z\hbar}\sum _\beta \mathrm{} ^\mathrm{'}\sum_\alpha\left[\hat{H}_0,\hat{z}\right]_{\alpha\beta}G^<_{\beta\alpha},
\end{equation}
where $\hat{H}_0$ is the Hamiltonian of an electron in the superlattice potential, $\hat{z}$ its coordinate operator, 
$L_z$ is the period of the structure, and the summation over $\beta$ takes place over the states of one period only (called the central period), which is emphasized by the prime in the summation. In view of the  approximations  introduced to limit the range of the Hamiltonian and the Green's functions, the summation over $\alpha$ then takes place over the states in the central period and its few nearest neighbors only.

The current given by the expression (\ref{eq:chap9current}) was interpreted in Ref.~\onlinecite{prb73-245320} to be entirely coherent, where the scattering events only redistribute the carriers in energy domain. Following that interpretation, the origin of all resonances, presented in the sections that follow, can be explained in terms of oscillations of coherence between ground states of neighboring periods, when the external field is varied. However, such an interpretation would not give an insight into the origin of the mentioned coherence oscillations. It has also been pointed out in Ref.~\onlinecite{prb73-245320} that in the basis of Wannier-Stark states coherences are created by scattering.
% , leading to a well known picture of scattering induced transport between Wannier-Stark ladder of states.
A very useful view of how coherences are created by scattering comes from the interpretation of the Keldysh relation. The interpretation in the time domain~\cite{pr357-1} considers $\Sigma^<$ as a scattering event, which is then propagated by $G^R$ and $G^A$ to a moment of time when coherence $G^<$ is observed. Following a similar interpretation that can be given in the energy domain and the fact that current is entirely determined by coherences, one can determine the origin of current in the structure, as follows.
In the case when $\alpha=\gamma$, $\delta=\beta$ and $\gamma\ne\delta$, the contribution to current from $G^<_{\alpha\beta}(E)$ originates from a scattering event (represented by $\Sigma^<_{\gamma\delta}(E)$) creating coherence at energy $E$, which will be observed only if there is available density of states (information about which is contained in $G^R_{\alpha\gamma}(E)$ and $G^A_{\delta\beta}(E)$) at that energy. On the other hand, when $\alpha=\gamma$, $\delta\ne\beta$ and $\gamma=\delta$, the current originates from a coherent propagation $G^A_{\delta\beta}(E)$, which will be observed providing there are carriers scattered into $\gamma=\delta$ (the term $\Sigma^<_{\gamma\delta}(E)$) and available density of states (the term $G^R_{\alpha\gamma}(E)$). The same interpretation of a coherent origin of current can be given in the case $\alpha\ne\gamma$, $\delta=\beta$ and $\gamma=\delta$. Other cases where the current originates from a combination of scattering and coherent propagation are also possible, but it is expected that these, being higher order processes, give a much smaller contribution. The results of the calculation presented here will indeed show that this is the case.

The current-field characteristics were calculated for a quantum dot array consisting of quantum dots whose dimensions are given in Sec.~\ref{Sec:chap9SCBA} for several different values of the period of the structure, using the same value of LO phonon linewidth $\Gamma$. The value chosen corresponds to a phonon lifetime of 5~ps which is within the range of the experimentally observed lifetimes.~\cite{prl44-1505,sse31-419} We have also verified that variations of $\Gamma$ in this range do not yield any significant qualitative differences in the results presented, although they of course give certain quantitative differences. As already emphasized in Sec.~\ref{Sec:chap9theory}, the calculation considers the Hamiltonian matrix elements, the Green's functions and the self-energies only among states with $|\Delta n|\le K$. This approximation is motivated by the fact that the Hamiltonian of the electron system and the Hamiltonian of the interaction with phonons are both short-ranged in a localized basis used. However, one cannot \textit{a priori} know whether the Green's functions will be short-ranged, as well. The calculations are therefore performed by increasing the value of $K$ until convergence is achieved (the values of $K$ that yield convergent results are reported for each calculation). The fact that convergence is achieved gives \textit{a posteriori} justification of the assumption of short-ranged nature of Green's functions. If Green's functions were long-ranged, the convergence would not have been achieved. 

\begin{figure}[htbp]
\vspace{1cm}
\centering
 \epsfig{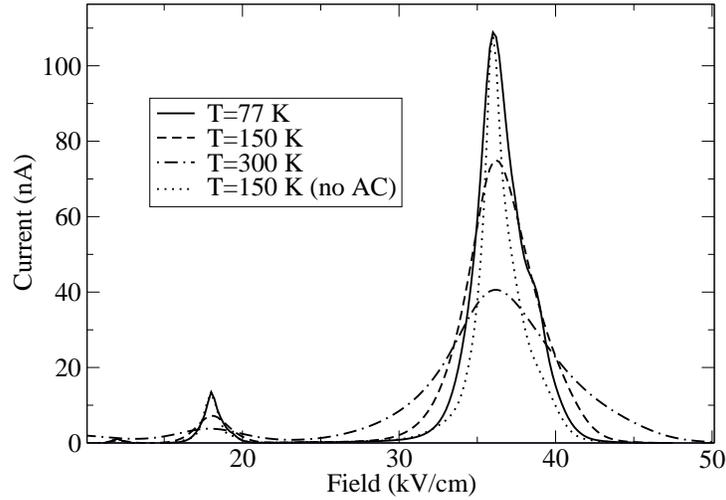}
 \caption{Current -- electric field characteristics of a quantum dot superlattice at temperatures $T=77\mrd\mathrm{K}$ (full line), $T=150\mrd\mathrm{K}$  with the influence of acoustic phonons (dashed line) and without it (dotted line), and $T=300\mrd\mathrm{K}$ (dashed-dotted line) when the period is $L_z=10\mrd\mathrm{nm}$.}
  \label{fig:chap9lz10}
\end{figure}

The results of the calculation for different temperatures when the period is equal to $L_z=10\mrd\mathrm{nm}$ are given in Fig.~\ref{fig:chap9lz10}. It was necessary to take  $K=2$ in the calculation to obtain convergent results. Self-energies due to the interaction with LO and acoustic phonons were both included in the calculation. Only the states originating from the ground miniband were considered at $T=77\mrd\mathrm{K}$ and $T=150\mrd\mathrm{K}$ since these are the only ones that are significantly populated then, while it was necessary to include a pair of first excited states at $T=300\mrd\mathrm{K}$. 

In order to understand the role of acoustic phonons, calculations have been performed where their contribution was excluded. The result at $T=150\mrd\mathrm{K}$ is shown by dotted line in Fig.~\ref{fig:chap9lz10}. Acoustic phonons have an energy which is too small to cause peaks in the characteristics, however they broaden the peaks caused by LO phonons and therefore play a certain role. Such a conclusion is fully in line with the results of Sec.~\ref{Sec:chap9SCBA} where it has been also shown that acoustic phonons cause significant broadening in the density of states.

\section{The main current peak}\label{Sec:chap9mainpeak}

The main peak arises when the potential drop over one period $V_F$ is equal to the LO phonon energy $E_\mathrm{LO}$. The second peak appears at $V_F=\frac{1}{2}E_{\mathrm{LO}}$ at all temperatures, while there is also a third peak at $V_F=\frac{1}{3}E_{\mathrm{LO}}$ present at lower temperatures. The origin of these resonances, as well as the nature of the electron transport at resonances will be investigated in what follows. The resonances predicted in the results reported here are in full analogy with Stark-cyclotron-phonon resonances or Stark-magneto-phonon resonances in the case of quantum well superlattices in a magnetic field.

The results of the calculation were transformed to a Wannier-Stark basis which is more useful for the physical interpretation of the results.
The focus will be given on the case of low temperatures when only the ground state is occupied. 
In that case, one can show using the properties of translational invariance and the identity $G^<_{\alpha\beta}=-G^<_{\beta\alpha}\hspace{0cm}^*$ that in the Wannier-Stark basis, the expression (\ref{eq:chap9current}) reduces to
\begin{equation}\label{eq:chap9current1}
I=\frac{e^2F}{\hbar}\sum_{\alpha>0}\alpha\cdot 2\mathrm{Re}\left(z_{0\alpha}G^<_{\alpha 0}\right).
\end{equation}
For $L_z=10\mrd\mathrm{nm}$, the current is entirely determined by the  $\alpha=1$ term, i.e. by the coherence between two ground states of neighboring periods $G^<_{10}(E)$. In order to understand the origin of the current, one therefore has to investigate the origin of this coherence. It should be mentioned that the fact that $G^<_{10}(E)$ determines the current does not necessarily imply that  the $K=1$ approximation is sufficient. Indeed, in this particular case, convergent results are obtained with $K=2$.

\begin{figure}[htbp]
\vspace{1cm}
\centering
 \epsfig{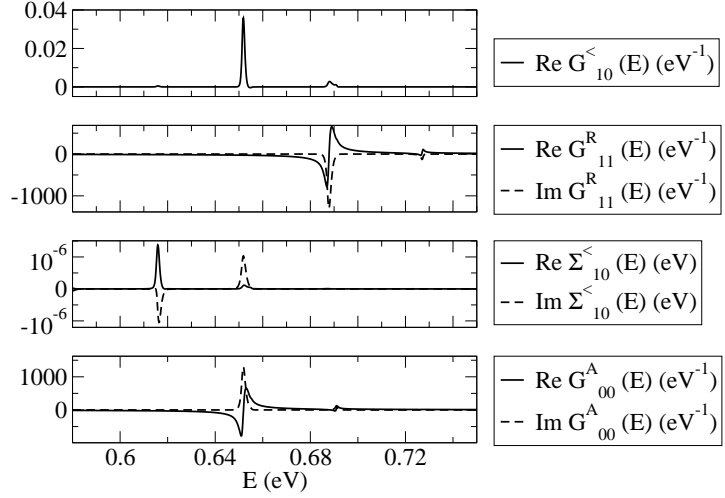}
  \caption{Green's functions and self-energies of the dominant contribution $G^R_{11}(E)\Sigma^<_{10}(E)G^A_{00}(E)$ to coherence $G^<_{10}(E)$, and hence the current, at LO phonon resonance ($V_F=E_{\mathrm{LO}}$) when the period is $L_z=10\mrd\mathrm{nm}$.}
   \label{fig:chap9figLO}
\end{figure}

\begin{figure}[htbp]
\vspace{1cm}
\centering
 \epsfig{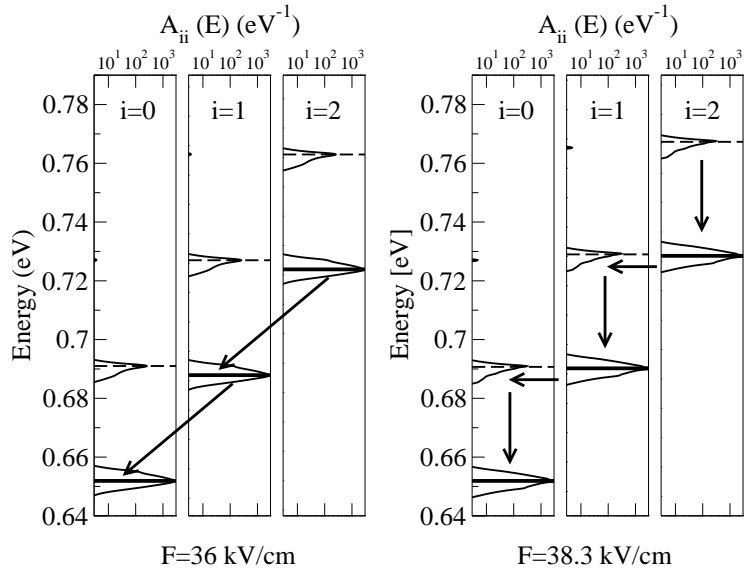}
 \caption{Schematic view of current transport at the field of $F=36\mrd\mathrm{kV/cm}$ corresponding to LO phonon resonance $V_F=E_{\mathrm{LO}}$ (left) and at $F=38.3\mrd\mathrm{kV/cm}$ corresponding to resonance of the phonon replica with the ground state of neighboring period (right). Corresponding density of states given by the spectral function $A_{ii}(E)=-2\mathrm{Im}G^R_{ii}(E)$ presented in logarithmic scale, is shown for each state. The dominant current transport channel in both cases is marked by arrows.}
  \label{fig:chap9sketch}
\end{figure}

The dominant contribution to $G^<_{10}(E)$ when $V_F=E_{\mathrm{LO}}$ comes from the $G^R_{11}(E)\Sigma^<_{10}(E)G^A_{00}(E)$ term in Keldysh relation. The corresponding Green's functions and self-energies are presented in Fig.~\ref{fig:chap9figLO}. $G^<_{10}(E)$ exhibits a maximum at the energy of level 0, originating from the maxima of the scattering $\Sigma^<_{10}(E)$ term and the $G^A_{00}(E)$ term.
In view of the interpretation of the Keldysh relation presented, the origin of the current at this value of the field is LO phonon scattering from level 1 to level 0, represented by the $\Sigma^<_{10}(E)$ term. By expressing $G^<_{10}$ in the energy domain as 
\begin{equation}
G^<_{10}=\frac{1}{2\pi}\int\mathrm{d}EG^<_{10}(E)
\end{equation}
and substituting into~(\ref{eq:chap9current1}) one can also spectrally resolve the current flow between periods 1 and 0. The maximum of the spectrally resolved current appears at the energy of the ground state of period 0, confirming the fact that the current flows into level 0, as demonstrated in the left part of Fig.~\ref{fig:chap9sketch}. 

\begin{figure}[htbp]
\vspace{1cm}
\centering
 \epsfig{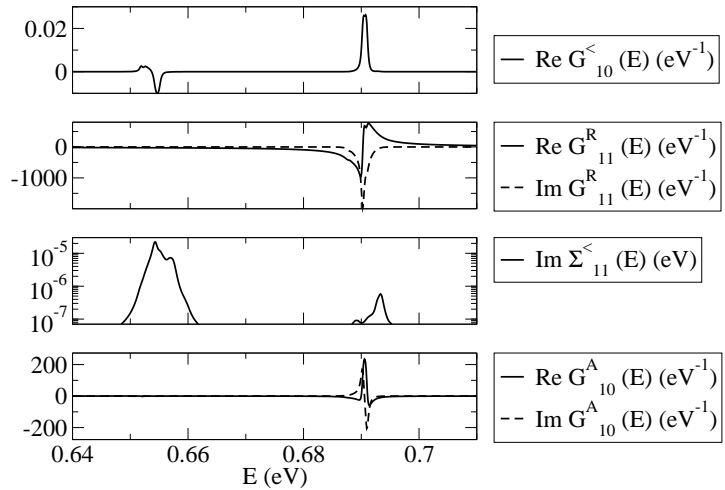}
 \caption{Green's functions and self-energies of the dominant contribution $ G^R_{11}(E)\Sigma^<_{11}(E)G^A_{10}(E)$ to coherence $G^<_{10}(E)$, at a field of $F=38.3\mrd\mathrm{kV/cm}$, corresponding to resonance of phonon replica with the state of the neighboring period. The period of the structure is $L_z=10\mathrm{nm}$.}
  \label{fig:chap9figPS}
\end{figure}

When one increases the electric field, the scattering $\Sigma^<_{10}(E)$ term decreases as 1 and 0 are no longer set to an LO phonon resonance. One should note that first phonon replica in the spectral density of states $A_{00}(E)=-2\mathrm{Im}G^R_{00}(E)$ (see Fig.~\ref{fig:chap9sketch}) is separated from the main maximum by an energy larger than $E_\mathrm{LO}$, as a consequence of the polaron shift, as demonstrated in Sec.~\ref{Sec:chap9SCBA}. Consequently, the resonance between the level 1 and phonon replica of level 0 occurs at a higher field, which in this particular case corresponds to a potential drop per period of $V_F=38.3\mrd\mathrm{meV}$ rather than $V_F=E_{\mathrm{LO}}=36\mrd\mathrm{meV}$. Around this resonance, the nature of the electron transport is significantly different than at an LO phonon resonance. The dominant contribution to coherence $G^<_{10}(E)$, shown in Fig.~\ref{fig:chap9figPS}, now comes from the $ G^R_{11}(E)\Sigma^<_{11}(E)G^A_{10}(E)$ term. Therefore, the current originates from coherent propagation represented by $G^A_{10}(E)$, which now exhibits a pronounced maximum at the energy of level 1. The coherence $G^<_{10}(E)$, and hence the current exhibit a maximum at the energy of level 1 (see Fig.~\ref{fig:chap9figPS}), confirming the interpretation that the transport channel at this value of the field is coherent tunneling to phonon replica, as shown schematically by horizontal arrows in the right part of Fig.~\ref{fig:chap9sketch}.

\begin{figure}[htbp]
\vspace{1cm}
\centering
 \epsfig{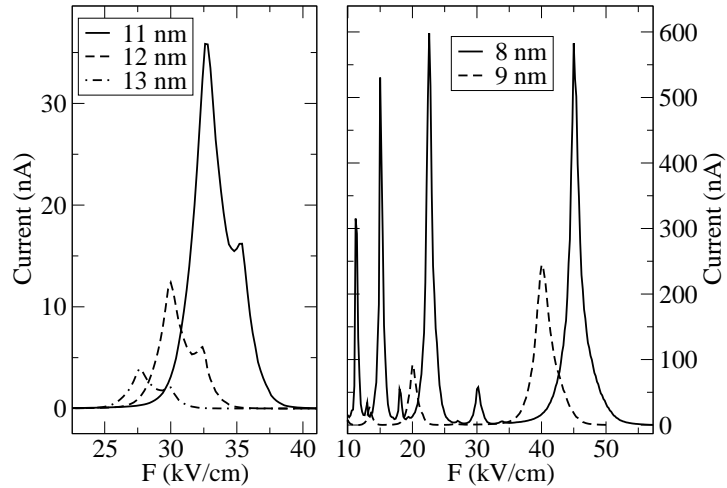}
 \caption{Current -- field characteristics for different values of the period, when the temperature is $T=77\mrd\mathrm{K}$. Convergent results are obtained with $K=4$ when $L_z=8\mrd\mathrm{nm}$, $K=3$ when $L_z=9\mrd\mathrm{nm}$, and $K=1$ when $L_z\ge 11\mrd\mathrm{nm}$.}
  \label{fig:chap9JFdif}
\end{figure}

At the period length of 10~nm and smaller, the two resonances cannot be distinguished as their separation is smaller than their width. However, at a larger value of the period when the linewidth decreases, the peaks become distinguishable, as shown in the left part of Fig.~\ref{fig:chap9JFdif}. 

From the previous discussion, it follows that the origin of the doublet structure is the fact that polaron replica of the ground state is at an energy different than $E_0+E_\mathrm{LO}$.
The doublet structure of the current peak is therefore a transport signature of polaron effects, where the separation between the peaks in the doublet is a measure of the electron -- phonon interaction strength. Polaron effects in self-assembled quantum dots have so far been evidenced by optical means only in the intraband magneto-optical absorption spectrum,~\cite{prl83-4152,PhysRevB.56.1516} magneto-photoluminescence spectrum~\cite{prb73-075320} and by Raman scattering.~\cite{prb73-233311} The results presented here therefore suggest a new physical effect: the manifestation of polaron effects in electron transport.

\section{Other resonances}\label{Sec:chap9otherres}

\begin{figure}[htbp]
\vspace{1cm}
\centering
 \epsfig{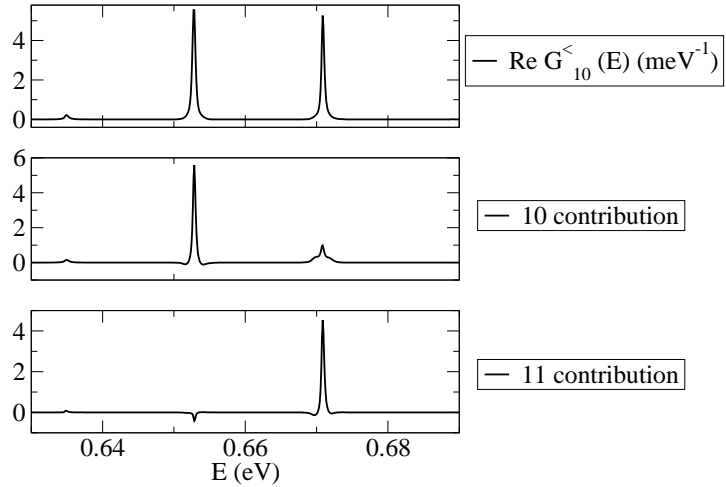}
  \caption{The coherence $G^<_{10}(E)$, at $V_F=\frac{1}{2}E_{\mathrm{LO}}$ and $L_z=10\mrd\mathrm{nm}$, as well as dominant contributions to it $G^R_{11}(E)\Sigma^<_{11}(E)G^A_{10}(E)$ (termed as 11 contribution) and 
  $G^R_{11}(E)\Sigma^<_{10}(E)G^A_{00}(E)$ (termed as 10 contribution).}
   \label{fig:chap9figDR}
\end{figure}

\begin{figure}[htbp]
\vspace{1cm}
\centering
 \epsfig{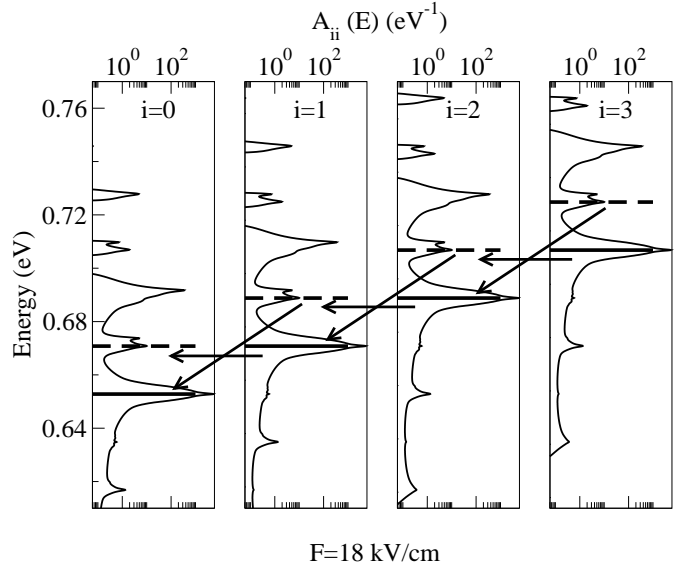}
 \caption{Schematic view of current transport at the period length $L_z=10\mrd\mathrm{nm}$ and the field of $F=18\mrd\mathrm{kV/cm}$ corresponding to $V_F=\frac{1}{2}E_{\mathrm{LO}}$. Corresponding density of states given by the spectral function $A_{ii}(E)=-2\mathrm{Im}G^R_{ii}(E)$, presented in logarithmic scale, is shown for each state. The dominant current transport channels are marked by arrows.}
  \label{fig:chap9sketchDR}
\end{figure}

The discussion will now be concentrated on a peak appearing at $V_F=\frac{1}{2}E_{\mathrm{LO}}$. While one might expect that the $\alpha=2$ term in (\ref{eq:chap9current1}) is of importance here, this is not the case, i.e. $G^<_{10}(E)$ mainly determines the current, as already mentioned. The dominant contribution to it comes in this case both from the scattering $G^R_{11}(E)\Sigma^<_{10}(E)G^A_{00}(E)$ term and the coherent $G^R_{11}(E)\Sigma^<_{11}(E)G^A_{10}(E)$ term, where each of these becomes dominant at an appropriate energy, as demonstrated in Fig.~\ref{fig:chap9figDR}.
In order to understand such behavior, one should note that the peaks in the spectral function $A_{ii}(E)$ appear not only at the energy of state $i$ and its phonon replica, but also at the energies of other states and their replicas. This is a consequence of the fact that in the presence of an interaction the Wannier-Stark states are no longer the eigenstates of the Hamiltonian of the system. The interaction then couples different Wannier-Stark states, with peaks appearing in the density of states as a consequence. Resonances in transport then appear when the peaks in the density of states of different periods overlap. In this particular case, the peak at $\frac{1}{2}E_\mathrm{LO}$ above the ground state of period $i$, being a consequence of LO phonon coupling with the ground state of period $(i-1)$, becomes resonant with the ground state of period $(i+1)$.
The scattering contribution to current between periods 1 and 0 therefore comes from the LO phonon scattering from the density of states at $\frac{1}{2}E_\mathrm{LO}$ above the ground state of period 1 to ground state of period 0. On the other hand, the coherent contribution comes from tunneling from the ground state of period 1 to the density of states at $\frac{1}{2}E_\mathrm{LO}$ above the ground state of period 0. These two contributions are schematically illustrated in Fig.~\ref{fig:chap9sketchDR}. One therefore sees that the transport between the ground state of period $i$ and the ground state of period $i-2$ which are at LO phonon resonance takes place by a sequence of two events: tunneling event represented by horizontal arrows in Fig.~\ref{fig:chap9sketchDR} and scattering event represented by diagonal arrows. As the two types of events follow each other, they yield nearly the same contributions to $G^<_{10}$, as demonstrated in Fig.~\ref{fig:chap9figDR}.

When the dots in a superlattice are closer, additional peaks in the transport appear. For example when the period is 8~nm, clearly visible peaks at $V_F=E_\mathrm{LO}$, $\frac{1}{2}E_\mathrm{LO}$, $\frac{1}{3}E_\mathrm{LO}$, $\frac{1}{4}E_\mathrm{LO}$, and even $\frac{2}{3}E_\mathrm{LO}$ and $\frac{2}{5}E_\mathrm{LO}$, can be seen in Fig.~\ref{fig:chap9JFdif}. 

The results obtained confirm the necessity of employing a model where coherent and polaron effects are fully taken into account, such as in the nonequilibrium Green's functions formalism. A semiclassical Boltzmann equations model would not be able to predict the doublet structure of the main peak and it would yield peaks in the current only at $V_F=E_\mathrm{LO}/n$ (where $n$ is an integer)  when the transition rates are evaluated within first order perturbation theory, while higher orders of perturbation theory would be necessary for the other peaks.

\section{Nonuniformities of the quantum dot ensemble}\label{sec:chap9secNON}

The discussion so far has addressed ideal periodic quantum dot arrays. However, real quantum dot ensembles are nonuniform and in a real experiment, one cannot expect to obtain the results predicted by the theory assuming ideal periodicity. On the other hand, the inclusion of quantum dot nonuniformity in the theory requires detailed information about the quantum dot size distribution and is obviously sample dependent. In order to estimate the influence of nonuniformities, additional self-energies were included in the theory according to the following approach.

Let $V$ be the additional potential due to the difference between the potential of a real ensemble of dots and an ideal dot superlattice. Within the SCBA, the contribution to self-energy from this potential is given by
\begin{equation}
\Sigma^{<,R}_{\alpha\beta}(E)=\sum_{\gamma\delta}\langle V_{\alpha\gamma}V_{\delta\beta}\rangle G^{<,R}_{\gamma\delta}(E).
\end{equation}
The average value $\langle V_{\alpha\gamma}V_{\delta\beta}\rangle$ contains  information about the quantum dot nonuniformities, and it should be in principle evaluated from the information provided by the experimental dot size distribution, which is sample dependent. For the purpose of an estimate which could be utilized regardless of the details of the dot distribution, a simple phenomenological approach is adopted here. It is assumed that $\langle V_{\alpha\gamma}V_{\delta\beta}\rangle=U^2$ when states $\alpha$, $\beta$, $\gamma$ and $\delta$ belong to the same period, and $\langle V_{\alpha\gamma}V_{\delta\beta}\rangle=0$ otherwise, where $U$ is a constant roughly representing the standard deviation of the position of quantum dot energy levels due to nonuniformities. This approach therefore assumes zero overlap of the matrix elements of the $V$-operator between the states of different periods, which is a reasonable assumption. Additionally, it assumes there is no correlation between the influence of nonuniformities on the states of different periods. Finally, the most severe assumption which makes this approach only an estimate is that $U$ is independent of $\alpha$, $\beta$, $\gamma$ and $\delta$, when these belong to the same period. However, when the transport takes place through ground states only, and therefore only one state per period is involved, as is the case here, this approximation becomes justified as well.

\begin{figure}[htbp]
\vspace{1cm}
\centering
 \epsfig{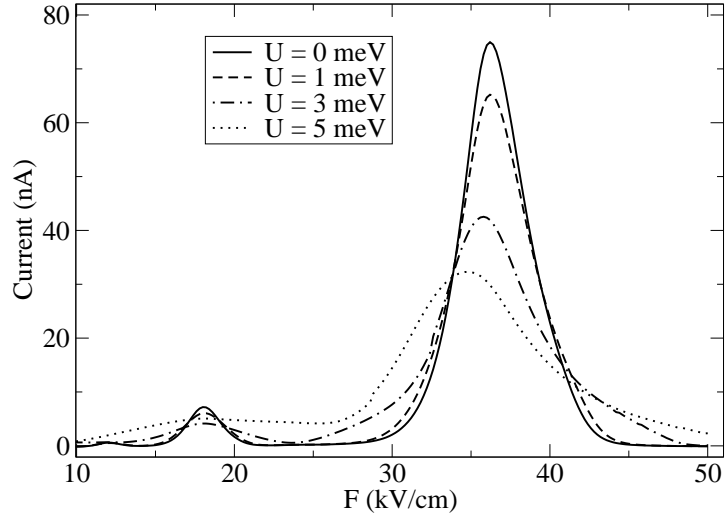}
 \caption{Current -- field characteristics for several values of phenomenological nonuniformity parameter $U$ at $T=150\mrd\mathrm{K}$, when the period of the structure is $L_z=10\mrd\mathrm{nm}$.}
  \label{fig:chap9figIFR}
\end{figure}

The current -- field characteristics for several different values of $U$ at $T=150\mrd\mathrm{K}$ and $L_z=10\mrd\mathrm{nm}$ are presented in Fig.~\ref{fig:chap9figIFR}. As expected, an increase in $U$ leads to broadening of the current peaks, with weaker peaks eventually vanishing. The main peak however, although broadened, remains clearly distinguishable.

\begin{figure}[htbp]
\vspace{1cm}
\centering
 \epsfig{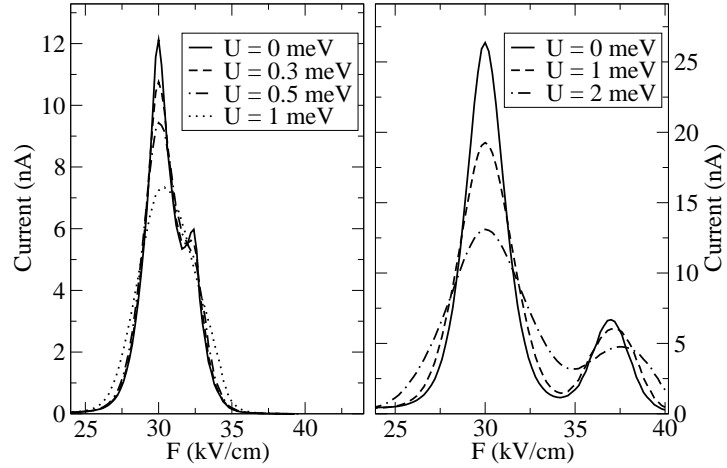}
 \caption{Current -- field characteristics for several values of the phenomenological nonuniformity parameter $U$ at $T=77\mrd\mathrm{K}$, when the period of the structure is $L_z=12\mrd\mathrm{nm}$ (left panel). The same result when electron -- LO phonon interaction Hamiltonian is multiplied by a factor of 2 is shown in the right panel.}
  \label{fig:chap9figIFR12}
\end{figure}

It is also interesting to estimate how nonuniformity affects the doublet structure of the main current peak. The $I$--$F$ curve at $T=77\mrd\mathrm{K}$ with different nonuniformity parameters is presented in the left panel of Fig.~\ref{fig:chap9figIFR12} for the structure with the period length of $L_z=12\mrd\mathrm{nm}$. One can conclude that already a weak nonuniformity of $U\sim 0.5\mrd\mathrm{meV}$ broadens the stronger peak of the doublet in such a way that the weaker peak vanishes. Therefore, in the InAs/GaAs material system the doublet structure could be observable only in extremely high uniform samples. On the other hand, InAs/GaAs is a system with weak polar coupling and one can expect a more favorable situation in systems with stronger coupling. The right panel of Fig.~\ref{fig:chap9figIFR12} presents the current -- field curve when the LO phonon interaction strength is multiplied by a factor of 2. In this case, the doublet structure remains observable even for nonuniformities of several meV. Therefore, although InAs/GaAs is not the most appropriate system for observing the signature of polaronic effects in electron transport, one can expect the effect to be observable in other systems.

\section{Conclusion}

Transport through bound states in periodic arrays of closely stacked quantum dots was analyzed. An appropriate theoretical framework based on nonequilibrium Green's functions formalism was developed and applied to calculate the current -- field characteristics. As expected, the current exhibits a strong peak when the potential drop over a period is equal to the LO phonon energy. The nature of charge transport at this resonance was analyzed in detail. It was found that at low temperatures the peak exhibits a doublet structure with one peak originating from LO phonon scattering between states of neighboring periods and the other one from resonant tunneling to a phonon replica of the state of the neighboring period. Therefore the doublet structure can be considered to be a transport signature of polaron effects. The nonuniformities of the quantum dot ensemble act to suppress the weaker peaks, while the main peak remains present. 

\bibliographystyle{h-physrev3}

\begin{thebibliography}{10}

\bibitem{phe14-115}
A.~J. Nozik,
\newblock Physica E {\bf 14}, 115 (2002).

\bibitem{prl97-247701}
A.~Marti, E.~Antolin, C.~R. Stanley, C.~D. Farmer, N.~Lopez, P.~Diaz,
  E.~Canovas, P.~G. Linares, and A.~Luque,
\newblock Phys. Rev. Lett. {\bf 97}, 247701 (2006).

\bibitem{apl82-415}
A.~A. Balandin and O.~L. Lazarenkova,
\newblock Appl. Phys. Lett. {\bf 82}, 415 (2003).

\bibitem{sci297-2229}
T.~C. Harman, P.~J. Taylor, M.~P. Walsh, and B.~E. LaForge,
\newblock Science {\bf 297}, 2229 (2002).

\bibitem{IEEEJQE33-1170}
N.~S. Wingren and C.~A. Stafford,
\newblock IEEE J. Quantum Electron. {\bf 33}, 1170 (1997).

\bibitem{IEEEJSTQE6-491}
C.-F. Hsu, J.-S. O, P.~Zory, and D.~Botez,
\newblock IEEE J. Select. Topics Quantum Electron. {\bf 6}, 491 (2000).

\bibitem{pss(a)202-987}
I.~A. Dmitriev and R.~A. Suris,
\newblock Phys. Status Solidi A {\bf 202}, 987 (2005).

\bibitem{PhysRevB.43.9336}
R.~Ferreira,
\newblock Phys. Rev. B {\bf 43}, 9336 (1991).

\bibitem{PhysRevB.59.8152}
V.~V. Bryksin and P.~Kleinert,
\newblock Phys. Rev. B {\bf 59}, 8152 (1999).

\bibitem{PhysRevB.53.7937}
N.~H. Shon and H.~N. Nazareno,
\newblock Phys. Rev. B {\bf 53}, 7937 (1996).

\bibitem{PhysRevB.56.15827}
P.~Kleinert and V.~V. Bryksin,
\newblock Phys. Rev. B {\bf 56}, 15827 (1997).

\bibitem{PhysRevLett.76.3618}
L.~Canali, M.~Lazzarino, L.~Sorba, and F.~Beltram,
\newblock Phys. Rev. Lett. {\bf 76}, 3618 (1996).

\bibitem{apl88-052111}
D.~Fowler, A.~Patane, A.~Ignatov, L.~Eaves, M.~Henini, N.~Mori, D.~K. Maude,
  and R.~Airey,
\newblock Appl. Phys. Lett. {\bf 88}, 052111 (2006).

\bibitem{prb74-121306}
R.~S. Deacon, R.~J. Nicholas, and P.~A. Shields,
\newblock Phys. Rev. B {\bf 74}, 121306(R) (2006).

\bibitem{prb73-115338}
C.~Gnodtke, G.~Kieszlich, E.~Scholl, and A.~Wacker,
\newblock Phys. Rev. B {\bf 73}, 115338 (2006).

\bibitem{prb66-085311}
D.~M.-T. Kuo and Y.~C. Chang,
\newblock Phys. Rev. B {\bf 66}, 085311 (2002).

\bibitem{Haug}
H.~Haug and A.-P. Jauho,
\newblock {\em {Quantum kinetics in transport and optics of semiconductors}}
  (Springer, Berlin, 1996).

\bibitem{pr357-1}
A.~Wacker,
\newblock Phys. Rep. {\bf 357}, 1 (2002).

\bibitem{sst21-1098}
N.~Vukmirovi\'c, \v{Z}. Ga\v{c}evi\'c, Z.~Ikoni\'c, D.~Indjin, P.~Harrison, and
  V.~Milanovi\'c,
\newblock Semicond.~Sci.~Technol. {\bf 21}, 1098 (2006).

\bibitem{Mahan}
G.~Mahan,
\newblock {\em Many-Particle Physics} (Kluwer Academic, 2000).

\bibitem{apl65-469}
C.-Y. Tsai, L.~F. Eastman, Y.-H. Lo, and C.-Y. Tsai,
\newblock Appl.~Phys.~Lett. {\bf 65}, 469 (1994).

\bibitem{prb45-8756}
L.~F. Register,
\newblock Phys.~Rev.~B {\bf 45}, 8756 (1992).

\bibitem{Harrison}
P.~Harrison,
\newblock {\em {Quantum Wells, Wires and Dots, 2$^{\mathrm{nd}}$ edition}}
  (John Wiley and Sons Ltd., Chichester, England, 2005).

\bibitem{ap236-1}
P.~Hyldgaard, S.~Hershfield, J.~H. Davies, and J.~W. Wilkins,
\newblock Ann. Phys. {\bf 236}, 1 (1994).

\bibitem{prb59-5069}
X.-Q. Li, H.~Nakayama, and Y.~Arakawa,
\newblock Phys.~Rev.~B {\bf 59}, 5069 (1999).

\bibitem{prl85-1516}
A.~V. Uskov, A.-P. Jauho, B.~Tromborg, J.~M\o{}rk, and R.~Lang,
\newblock Phys. Rev. Lett. {\bf 85}, 1516 (2000).

\bibitem{PhysRevB.66.245314}
S.-C. Lee and A.~Wacker,
\newblock Phys. Rev. B {\bf 66}, 245314 (2002).

\bibitem{prb71-125327}
J.~Seebeck, T.~R. Nielsen, P.~Gartner, and F.~Jahnke,
\newblock Phys. Rev. B {\bf 71}, 125327 (2005).

\bibitem{PhysRevB.62.7336}
T.~Stauber, R.~Zimmermann, and H.~Castella,
\newblock Phys. Rev. B {\bf 62}, 7336 (2000).

\bibitem{prb73-115303}
T.~Stauber and R.~Zimmermann,
\newblock Phys. Rev. B {\bf 73}, 115303 (2006).

\bibitem{prb73-245320}
S.-C. Lee, F.~Banit, M.~Woerner, and A.~Wacker,
\newblock Phys. Rev. B {\bf 73}, 245320 (2006).

\bibitem{prb72-125347}
R.~C. Iotti, E.~Ciancio, and F.~Rossi,
\newblock Phys. Rev. B {\bf 72}, 125347 (2005).

\bibitem{prl44-1505}
D.~von~der Linde, J.~Kuhl, and H.~Klingenberg,
\newblock Phys. Rev. Lett. {\bf 44}, 1505 (1980).

\bibitem{sse31-419}
J.~A. Kash and J.~C. Tsang,
\newblock Solid State Electron. {\bf 31}, 419 (1988).

\bibitem{prl83-4152}
S.~Hameau, Y.~Guldner, O.~Verzelen, R.~Ferreira, G.~Bastard, J.~Zeman,
  A.~Lema\^itre, and J.~M. G\'erard,
\newblock Phys. Rev. Lett. {\bf 83}, 4152 (1999).

\bibitem{PhysRevB.56.1516}
P.~A. Knipp, T.~L. Reinecke, A.~Lorke, M.~Fricke, and P.~M. Petroff,
\newblock Phys. Rev. B {\bf 56}, 1516 (1997).

\bibitem{prb73-075320}
V.~Preisler, T.~Grange, R.~Ferreira, L.~A. de~Vaulchier, Y.~Guldner, F.~J.
  Teran, M.~Potemski, and A.~Lemaitre,
\newblock Phys. Rev. B {\bf 73}, 075320 (2006).

\bibitem{prb73-233311}
B.~Aslan, H.~C. Liu, M.~Korkusinski, P.~Hawrylak, and D.~J. Lockwood,
\newblock Phys. Rev. B {\bf 73}, 233311 (2006).

\end{thebibliography}

\end{document}